\documentclass[floatfix,superscriptaddress,a4paper,
               showpacs,showkeys,nofootinbib,preprint]{revtex4}
\usepackage{epsfig}
\usepackage{latexsym}
\usepackage{xspace}
\usepackage{hyperref}
\usepackage{inputenc}
\usepackage{indentfirst}
\usepackage{enumerate}
\usepackage{color}

\usepackage{amsmath}
\usepackage{amssymb}
\usepackage[english]{babel}
\usepackage{url}
\newcommand{\eq}[1]{\begin{align} #1 \end{align}}

\newcommand{\eV}{\ensuremath{\mbox{e\kern-0.1em V}}\xspace}
\newcommand{\GeV}{\ensuremath{\mbox{Ge\kern-0.1em V}}\xspace}
\newcommand{\MeV}{\ensuremath{\mbox{Me\kern-0.1em V}}\xspace}
\newcommand{\GeVc}{\ensuremath{\mbox{Ge\kern-0.1em V}\!/\!c}\xspace}
\newcommand{\GeVcc}{\ensuremath{\mbox{Ge\kern-0.1em V}\!/\!c^2}\xspace}
\newcommand{\AGeV}{\ensuremath{A\,\mbox{Ge\kern-0.1em V}}\xspace}
\newcommand{\AGeVc}{\ensuremath{A\,\mbox{Ge\kern-0.1em V}\!/\!c}\xspace}
\newcommand{\MeVc}{\ensuremath{\mbox{Me\kern-0.1em V}/c}\xspace}

\newcommand{\dd}{\ensuremath{{\textrm d}}\xspace}
\newcommand{\dedx}{\ensuremath{\dd E\!/\!\dd x}\xspace}

\newcommand{\CernVM}{\textsc{Cern\-\kern-0.05emVM}\xspace}

\begin{document}

\title{Particle-Set Identification method to study multiplicity fluctuations }

\author{M. Gazdzicki}
\affiliation{Geothe-University Frankfurt am Main, Germany}
\affiliation{Jan Kochanowski University, Kielce, Poland}
\author{M. I. Gorenstein}
\affiliation{Bogolyubov Institute for Theoretical Physics, Kiev, Ukraine}
\affiliation{Frankfurt Institute for Advanced Studies, Frankfurt, Germany}
\author{M. Mackowiak-Pawlowska}
\affiliation{Faculty of Physics, Warsaw University of Technology, Warsaw, Poland}
\author{A. Rustamov}
\affiliation{GSI Helmholtzzentrum f{\"u}r Schwerionenforschung, Darmstadt, Germany}
\affiliation{National Nuclear Research Center, Baku, Azerbaijan}

\begin{abstract}
In this paper a new method of experimental data analysis, the Particle-Set Identification method,
is presented. The method allows to
reconstruct moments of multiplicity distribution of identified particles.
The difficulty the method copes with is due to incomplete particle identification --
a particle mass is frequently determined with a resolution which does not allow for a unique
determination of the particle type.
Within this method the
moments of order $k$ are calculated from mean multiplicities of $k$-particle sets of a given type.
The Particle-Set Identification method remains valid even in the case of correlations between
mass measurements for different particles.
This distinguishes it from the Identity method introduced by us previously to solve 
the problem of incomplete particle identification in studies of particle fluctuations.

\end{abstract}


\maketitle
\section{Introduction}
\label{sec:introduction}

Study of multiplicity fluctuations of identified particles produced in high energy nucleus-nucleus collisions is motivated by a number of important physics issues. They include
searching for the critical point of strongly interacting matter,
uncovering  properties of hadronization and
testing the validity of statistical models beyond mean multiplicities of identified hadrons,
see Refs.~\cite{Koch:2008ia, Gazdzicki:2015ska} for details.
Experimental measurements of multiplicity distributions of identified hadrons
are challenging because it is often impossible to identify a particle with sufficient precision. Typical tracking detectors, like
time projection chambers or silicon pixel detectors,
used in high energy physics, allow for a precise measurement of momenta
of charged particles and  sign of their electric charges.
In order to be able to distinguish between different
particle types (e.g. e$^+$, $\pi^+$, $K^+$, proton)
a determination of particle mass is necessary.
This is done indirectly by measuring  for each
particle the value of a special variable, e.g., specific energy loss, \dedx, time-of-flight or the Cerenkov radiation angle. 
In the following  this variable is referred to as the "mass" variable.
In 2011 the Identity method~\cite{Gazdzicki:2011xz}  was proposed as a tool to measure moments of
multiplicity distribution of identified particles, which circumvents  the experimental issue of
incomplete particle identification. The method was significantly
developed in Ref.~\cite{Gorenstein:2011hr} and extended in following
papers, see Refs.~\cite{Rustamov:2012bx, Pruneau:2017fim, Pruneau:2018glv, Mackowiak-Pawlowska:2017dma}.
Currently, the Identity method is used by several
experimental collaborations~\mbox{\cite{Anticic:2013htn,Mackowiak-Pawlowska:2013caa,Acharya:2017cpf,Rustamov:2017lio}}.

In this paper we present a novel approach, the Particle-Set (PSET) Identification method,
for reconstructing moments of multiplicity distribution of identified particles.
Hereinafter a PSET represents a set of
$k$-particles ($k=2,3,\ldots$) which is constructed from particles created in a collision.
For example, in the simplest non-trivial case of two particle types,
pions and kaons, three types of two-particle sets are possible: pion-pion, kaon-kaon and
pion-kaon. Mean multiplicities of different two-particle sets permit to calculate the second order moments of the joint multiplicity distribution of pions and kaons.
To find the third order (in general, $k^{th}$ order)  moments of the multiplicity distribution one needs mean multiplicities of three-particle (in general, $k$-particle) sets.
The latter are obtained from  fits to
the $k$-dimensional distribution of particle masses.
The PSET Identification method has a broader range of applicability than the Identity method.
Particularly, it does not assume that measurements of masses for different
particles are independent.

The paper is organized as follows.
Section~\ref{sec:pset} introduces the PSET Identification method.
For simplicity, the presentation in this section
is restricted to second moments of the multiplicity distribution
of two particle types, pions and kaons.
First, the problem is defined and the well-known method to calculate mean multiplicities of
identified hadrons is reviewed. Next an extension of this method, needed 
to calculate second moments, is presented.
In Sec.~\ref{sec:PSET-general}, a general formulation of the method is sketched. This allows to deal with an arbitrary number of particle types
as well as to calculate higher than second order moments. In Sec.~\ref{sec:simulation}, exploiting simple Monte Carlo simulations, the PSET Identification method is confronted with the Identity method.
Section~\ref{sec:summary} closes the paper.

\section{Introduction to PSET Identification method}
\label{sec:pset}


The resolution of the "mass" variable, denoted as $x$, is usually
poor -- often probabilities to register particles of different types in the same interval
of $x$ are comparable.
Consequently, it is impossible to identify particles individually
with a reasonable confidence level.
However, it can still be possible to extract information on average production properties
of  particles of a given type, such as mean multiplicity of single particles
or, in general, mean multiplicity of particle sets of a given type.

The resolution of the "mass" variable measurement for single particles of type $a$
is quantified by the probability density function (pdf) $f_a(x, p)$.
In general, the function depends also on the particle momentum vector $p$ 
which complicates the data analysis.
Detector properties, data calibration and analysis procedures determine the pdf $f_a(x, p)$.

When pairs of particles are of interest, the single particle pdf is replaced by a two-particle
probability density function, $f_{ab}\big( (x_1, p_1), (x_2, p_2) \big)$, where $a$ and $b$
are particle types of the first and second particle, respectively.
Single and two-particle pdfs are related as
\eq{ \label{eq:pdf21}
f_a( x_1, p_1 ) = \int dx_2~dp_2~f_{ab}\big( (x_1, p_1), (x_2, p_2) \big)~.
}
In general, the two-particle pdf cannot be represented as a product of the single particle pdfs: 
$f_{ab}~\big( (x_1,p_1), (x_2,p_2) \big)  \neq f_{a}(x_1,p_1) \cdot f_{b}(x_2,p_2)$.
The reason are correlations which are caused by detector and data calibration properties 
(e.g., a finite two track resolution and/or time dependent response functions) 
as well as by the analysis procedure (e.g., neglecting the $p$-dependence of the pdfs within finite momentum bins used
in  analyses).

For clarity, the PSET Identification method
is introduced  using the following simplifying assumptions:

\begin{enumerate}[(i)]

\item  There are only two particle types: positively charged pions and kaons which for
       simplicity are referred to as pions ($\pi$) and kaons ($K$).
       
\item  The single-particle and two-particle pdfs are independent of particle momenta, 
       i.e.,   $f_a(x,p)=f_a(x)$ and
       $f_{ab}\big( (x_1, p_1), (x_2, p_2) \big) = f_{ab}( x_1, x_2 )$,  for $a,b=\pi,K$.

\item  The pdfs $f_a(x)$ and $f_{ab}( x_1, x_2 )$  are
       assumed to be  known.
\end{enumerate}

The extension of this simple model to an arbitrary number of particle types
as well as to higher moments of  multiplicity distribution is discussed in the next section.

\subsection{Mean multiplicities of identified hadrons}
\label{sec:mm}

Let us assume that
an experiment measures particles produced in $M$ collisions (events).
The set of $x$ values measured for the $N_i$ particles in an event $i$
is denoted as
$X_i = \{x_1, x_2, ... , x_{N_i}\}$~.
The total number of particles measured in $M$ events is
\eq{\label{eq:NN}
\mathcal{N} = \sum_{i=1}^M N_i~,
}
and the mean particle multiplicity (sum of pions and kaons) can be calculated as
\eq{\label{eq:mN}
\langle N \rangle = \frac{\mathcal{N}}{M}~.
}
The full set of  $x$-measurements is denoted as $\mathcal{X} = \{x_1, x_2, ... , x_{\mathcal{N}}\}$~.


Statistical tools to extract mean multiplicities of identified hadrons and its momentum dependence
(single particle spectra) are presented and discussed in detail in Ref.~\cite{Gazdzicki:1994vj}.
Here the general solution~\cite{Gazdzicki:1994vj} is adapted
to the simple model introduced in this section.

The distribution of the sum of pions and kaons in the "mass" variable $x$ is given by
\eq{ \label{eq:rho}
\rho(x) \equiv \rho_\pi(x) + \rho_K(x) \equiv   \langle N_\pi \rangle \cdot f_\pi (x) + \langle N_K \rangle
\cdot f_K (x)~\equiv~\langle N\rangle \big( r_\pi\cdot f_\pi(x)+r_K\cdot f_K(x)\big),
}
where $f_\pi(x)$ and $f_K(x)$ are the known pdfs
of pions and kaons, respectively. Parameters $r_{\pi}$ and $r_{K}$ define what fraction of particles in a given set of events is either pion or kaon. By definition $r_\pi + r_K = 1$ and 
consequently, only one parameter, e.g. $r_K$, should be estimated from
the data $\mathcal{X}$.

Mean multiplicities of pions $\langle N_\pi \rangle$ and kaons $\langle N_K \rangle$ are unknown
and should be extracted from the experimental data  $\mathcal{X} = \{x_1, x_2, ... x_{\mathcal{N}} \}$~. They are equal to $\langle N_\pi \rangle = r_\pi \cdot \langle N \rangle $ and
$\langle N_K \rangle = r_K \cdot \langle N \rangle $ with $\langle N \rangle$ calculated from
the data as given by Eq.~\ref{eq:mN}.
Finally, we note that the function 
\eq{ \label{eq:F}
F(x)  =  r_\pi \cdot f_\pi (x) + r_K  \cdot f_K (x)
}
is the pdf for the system of pions and kaons.

\begin{figure}[htb]
\centering
 \includegraphics[width=0.45\linewidth,clip=true]{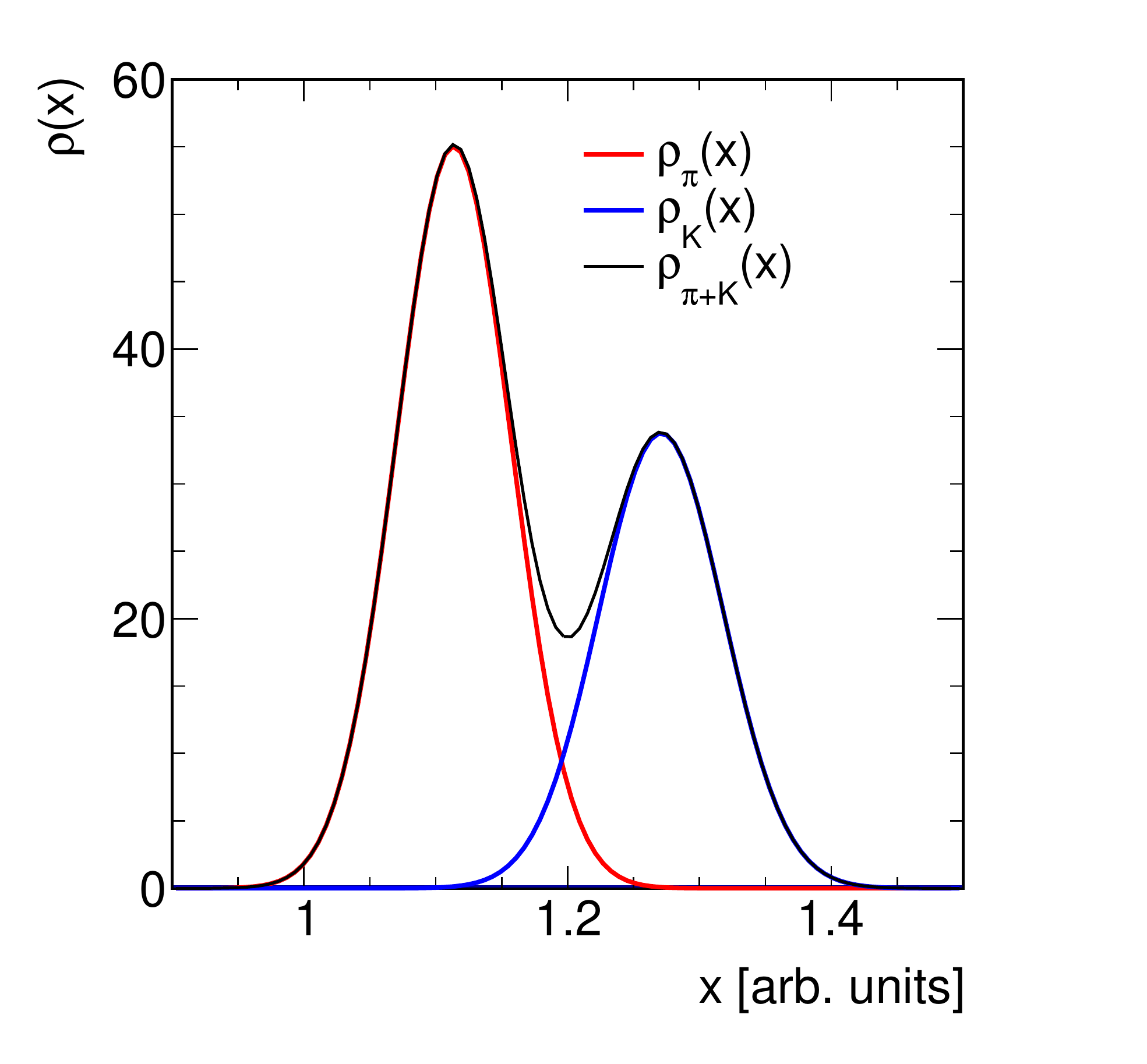}
 \includegraphics[width=0.54\linewidth,clip=true]{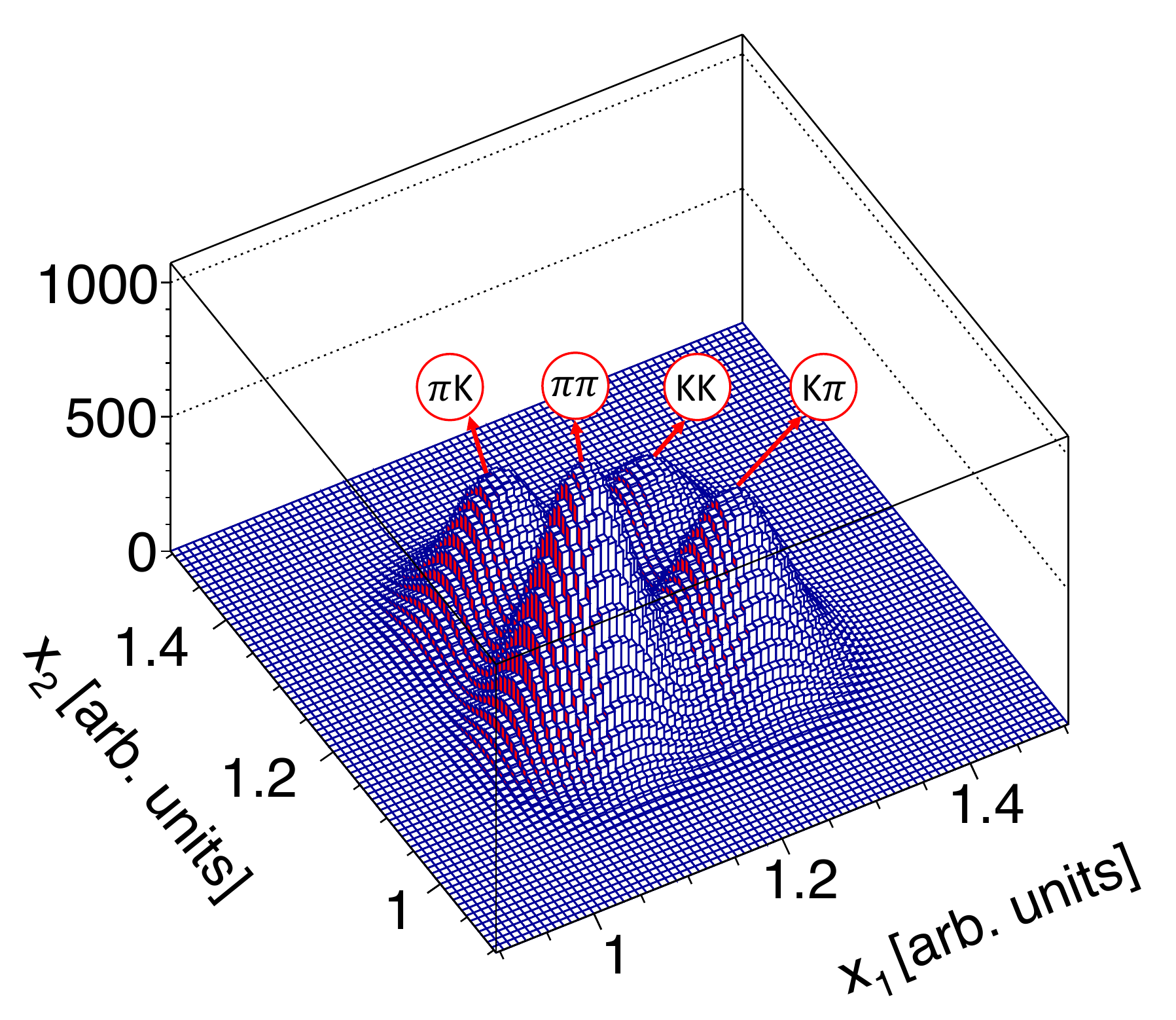}
 \caption{(Color online)
 \textit{Left panel:}
Example of the single-particle distribution $\rho(x)$ [Eq.~\ref{eq:rho}] 
for the system of pions and kaons.
\textit{Right panel:}
Example of the two-particle distribution $\rho(x_1,x_2)$ for the system of pions and kaons. 
The shown distribution is obtained by  assuming that $x_1$ and $x_2$ are independent,
i.e., $f_{ab}(x_1,x_2)=f_a(x_1)\cdot f_b(x_2)$, and 
pion and kaon multiplicities are uncorrelated and distributed according to
Poisson distributions.
The functions $f_\pi(x)$ and
$f_K(x)$, and  mean multiplicities of $N_\pi$ and $N_K$ correspond to those
shown in the left panel.
 }
\label{fig-1}
\end{figure}

An example distribution  of  the "mass" variable $x$, $\rho(x)$ (Eq.~\ref{eq:rho}),
for a system of pions and kaons is presented in Fig.~\ref{fig-1}, {\it left}.
The overlap of the $f_\pi(x)$ and $f_K(x)$ pdfs does not allow 
to identify a large fraction of particles with a high confidence level.

There are two commonly used parameter estimation methods
(see, e.g.,  Ref.~\cite{Cowan:2010bz}):
\begin{enumerate}[(i)]
    \item the least-squares method (LSM),
    \item the maximum likelihood method (MLM).
\end{enumerate}
The application of the LSM would require binning of the data points in $x$-space.
This has two important disadvantages:
the binning procedure causes partial loss of the experimental information, and
the binning necessary for the meaningful application of the LSM
(number of entries in each bin has to be large enough) may be impossible in the case of low
statistics data.

The above problems are not present in the MLM, as it allows to use the  unbinned data.
Therefore, in the following  we will briefly introduce the MLM and
the way it should be applied here.
Let us start from defining the likelihood function (LF):
\eq{ \label{eq:LF}
L(\mathcal{X}~|~r_K) = \prod_{j = 1}^{\mathcal{N}} F(x_j~|~r_K)
}
as a joint conditional probability of the measurements $\mathcal{X}$ at the fixed value of
the parameter $r_K$. Now, in the LF we treat the measurements $\mathcal{X}$ as
fixed values and the parameter $r_K$ as a variable. 
According to the maximum likelihood principle we should choose
a value of  $r_K=r_K^*$ which maximizes $L(\mathcal{X}~|~r_K)$.
It is usually more convenient to minimize the auxiliary function, $l(\mathcal{X}~|~r_K)$,
defined as :
\eq{ \label{eq:l}
l(\mathcal{X}~|~r_K) \equiv ~-\ln\big( L(\mathcal{X}~|~r_K) \big)=  -\sum_{j = 1}^{\mathcal{N}} 
\ln \big( F(x_j~|~r_K) \big)~.
}
The search for the value of $r_K^*$, which minimizes $l(\mathcal{X} | r_K)$,
has to be done using standard numerical minimization procedures~\cite{James:1975dr}.

To estimate the statistical uncertainty of $r_K^*$ one should use the
sub-sample~\cite{Tsao:2012} and/or bootstrap methods~\cite{Efron:1979} at the
level of events as independent data units.
It is important to stress that both methods take into account correlations
between measurements of the "mass" variable for different particles.
The goodness-of-fit tests are discussed in detail in Ref.~\cite{Gazdzicki:1994vj}.

\subsection{Mean multiplicity of particle pairs}
\label{sec:pairs}

In this subsection the PSET Identification method is considered for particle pairs.
Let us start with the observation that the mean multiplicity of
pairs of particles of a given type is directly related to the second moment
of its multiplicity distribution.
For example, the mean multiplicity of pion-pion pairs, $\langle N_{\pi\pi} \rangle$, is given as:
\eq{\label{eq:mNpp}
	\langle N_{\pi\pi} \rangle ~=~\frac{1}{2} \big\langle N_\pi\,(N_\pi -1) \big\rangle ~  = ~ \frac{\langle N_{\pi}^2 \rangle  - \langle N_{\pi} \rangle}{2}~.
}
When mean multiplicities  of pions, $\langle N_\pi \rangle$, and pion-pion pairs, $\langle N_{\pi\pi}\rangle$, are known,
the second moment of the pion multiplicity distribution  can be  calculated as
\eq{\label{eq:N2}
 	\langle N_{\pi}^2 \rangle ~=~ 2 \cdot \langle N_{\pi\pi} \rangle  + \langle N_{\pi} \rangle~.
}
In a similar way $\langle N_{K}^2 \rangle$ and $\langle N_\pi \cdot N_{K} \rangle$ are
calculated.
Thus the problem of measuring second moments of joint multiplicity distributions of identified particles is reduced to finding the mean multiplicities of identified particle pairs.

The experimental data on particle pairs in $M$ events is defined as follows.
Number of particle pairs in an event $i$ of multiplicity $N_i$ is
\eq{ \label{eq:event.pairs}
N_i^{(2)} = \frac{1}{2} N_i (N_i - 1)~.
}
The total number of pairs in all $M$ events is
\eq{ \label{eq:total.pairs}
\mathcal{N}^{(2)} = \sum_{i=1}^M  N_i^{(2)}~,
}
and the mean  multiplicity of all possible pairs can be calculated  as
\eq{\label{eq:mN2}
\langle N^{(2)} \rangle = \frac{\mathcal{N}^{(2)}}{M}~.
}
The full set of the pair data consists of $\mathcal{N}^{(2)}$ pairs:
\eq{ \label{eq:pairs.data}
\mathcal{X}^{(2)} = \{ (x_1, x_2)_1,  (x_1, x_2)_2,    \ldots , (x_1, x_2)_{\mathcal{N}^{(2)}} \}~.
}

The two-particle  mass distribution function $\rho(x_1,x_2)$ is a weighted sum
of two-dimensional  pdfs of identified pairs:
\eq{ \label{eq:rho2}
&	\rho(x_1,x_2)  \equiv \rho_{\pi\pi}(x_1,x_2) + \rho_{KK}(x_1,x_2) + \rho_{\pi K}(x_1,x_2) + \rho_{K \pi}(x_1,x_2)  \nonumber \\
    &	\equiv   \langle N_{\pi \pi} \rangle \cdot f_{\pi\pi}(x_1, x_2) +
	         \langle N_{K K} \rangle \cdot f_{KK}(x_1, x_2) +  \langle N_{\pi K} \rangle \cdot f_{\pi K}(x_1, x_2) +
	         \langle N_{K \pi} \rangle \cdot f_{K \pi}(x_1, x_2)~\nonumber \\
& \equiv \langle N^{(2)}\rangle \big( r_{\pi \pi} \cdot f_{\pi\pi}(x_1, x_2) +
	               r_{K K}  \cdot f_{KK}(x_1, x_2) +
	               r_{\pi K}  \cdot f_{\pi K}(x_1, x_2) +
	               r_{\pi K}  \cdot f_{K \pi}(x_1, x_2)~\big) , 
}
with $r_{\pi \pi} +  r_{K K} + 2 \cdot r_{\pi K}  = 1$~.
Consequently, only two parameters should be estimated from the experimental measurements
$\mathcal{X}^{(2)}$.
The two-dimensional probability density function for the system of pions and kaons reads:
\eq{ \label{eq:F2}
F(x_1,x_2 | r_{\pi\pi}, r_{KK} )  =  r_{\pi \pi} \cdot f_{\pi\pi}(x_1, x_2) +
	               r_{K K}  \cdot f_{KK}(x_1, x_2) +
	               r_{\pi K}  \cdot f_{\pi K}(x_1, x_2) +
	               r_{\pi K}  \cdot f_{K \pi}(x_1, x_2)
}
and using the MLM the auxiliary likelihood function:
\eq{ \label{eq:l2}
l(\mathcal{X}^{(2)}~|~r_{\pi \pi},r_{KK}) =  -\sum_{j = 1}^{\mathcal{N}^{(2)}} 
\ln \big( F\left((x_1,x_2)_j~|~r_{\pi\pi},r_{KK}\right) \big)~.
}
is minimized to determine the parameters $r_{\pi\pi}$ and $r_{KK}$.
Statistical uncertainties of the fitted parameter values,
r$_{\pi\pi}$ and r$_{KK}$ are to be estimated following the procedure described in the
previous subsection.
An example of the $\rho(x_1,x_2)$ function is presented in Fig.\ref{fig-1} {\it right}.


\section{PSET Identification method: towards general formulation}
\label{sec:PSET-general}

In this section 
the results presented in the previous section are extended to three particle types and
to  third  moments of the multiplicity distributions. Then, the extension to higher
number of particle types and higher  moments is obvious.

In the case of three particle types, for example, pions, kaons, and protons, the two-particle "mass" distribution function reads:
\eq{ \label{eq:rho2-3}
	\rho(x_1,x_2)
   & =   \langle N^{(2)}\rangle \big(~ r_{\pi \pi} \cdot f_{\pi \pi}(x_1,x_2) +
	         r_{K K}  \cdot f_{K K}(x_1,x_2) +
	         r_{pp} \cdot f_{pp}(x_1,x_2)   \nonumber \\
	  &     +  r_{\pi K}  \cdot f_{\pi K}(x_1,x_2) +
	          r_{K \pi}  \cdot f_{K \pi}(x_1,x_2)~ +
	         r_{\pi p}  \cdot f_{\pi p}(x_1,x_2) \nonumber \\
& +
	         r_{p \pi}  \cdot f_{p \pi}(x_1,x_2)+
	         r_{Kp} \cdot f_{K p}(x_1,x_2) +
	         r_{pK } \cdot f_{p K}(x_1,x_2)~\big)~.
}
From relations $r_{ab}=r_{ba}$
and $\sum_{a,b}r_{ab}=1$ it follows that only five parameters are independent. 
Their values can be found by minimizing the corresponding auxiliary likelihood function, 
as that in Eq.~(\ref{eq:l2}).
In a similar way expressions for more than three particle types can be obtained.

Let us now consider third moments of the multiplicity distribution 
for two particle types, pions and kaons.
To calculate them within the PSET Identification
method one should first extract from the data the mean multiplicity 
of identified three-particle sets.
Then the three-dimensional pdfs, $f_{abc}(x_1,x_2,x_3)$, have to be known. 
With these functions, the three-particle pdfs
$F(x_1,x_2,x_3~|~r )$ can be calculated. Here $r$ denotes a set of unknown independent
parameters. Their values 
should be fitted to the data on particle triplets
\eq{ \label{eq:tr.data}
\mathcal{X}^{(3)} = \{ (x_1, x_2,x_3)_1,  (x_1, x_2,x_3)_2,    \ldots ,  (x_1, x_2, x_3)_{\mathcal{N}^{(3)}} \}~.
}

Mean multiplicities $\langle N_{abc} \rangle =r_{abc}\langle N^{(3)}\rangle$ are straightforwardly connected
with the third order moments of the identified particle multiplicity distributions.
For example,
\eq{
\langle N_{\pi\pi\pi} \rangle & = \frac{\langle N_\pi(N_\pi -1)(N_\pi -2)\rangle }{3!}=\frac{1}{6}\,\left[\,\langle N_\pi^3\rangle
-3\langle N_\pi^2\rangle+2\langle N_\pi\rangle \, \right]~,\\
\langle N_{\pi\pi K} \rangle & = \frac{\langle N_\pi(N_\pi -1) N_K \rangle}{2}=\frac{1}{2}\,\left[\,\langle N_\pi^2 N_K \rangle
-\langle N_\pi N_K\rangle \, \right]~.
}
In this way the mean multiplicity of the identified three-particle set, 
$\langle N_{abc} \rangle$, is
obtained by a linear combination of third, second, and first moments of the identified
particle multiplicity distributions.
Similarly, third moments  of  the multiplicity distribution can be derived
from  a linear combination of mean multiplicities of single particles, particle pairs, 
and particle triplets.

The above procedures are straightforwardly extendable to $k$-particle sets and  moments
of order $k$ with $k>3$ for an arbitrarily large number of particle types. 
\section{Test on simulated data}
\label{sec:simulation}

\begin{figure}[htb]
\centering
\includegraphics[width=0.45\linewidth,clip=true]{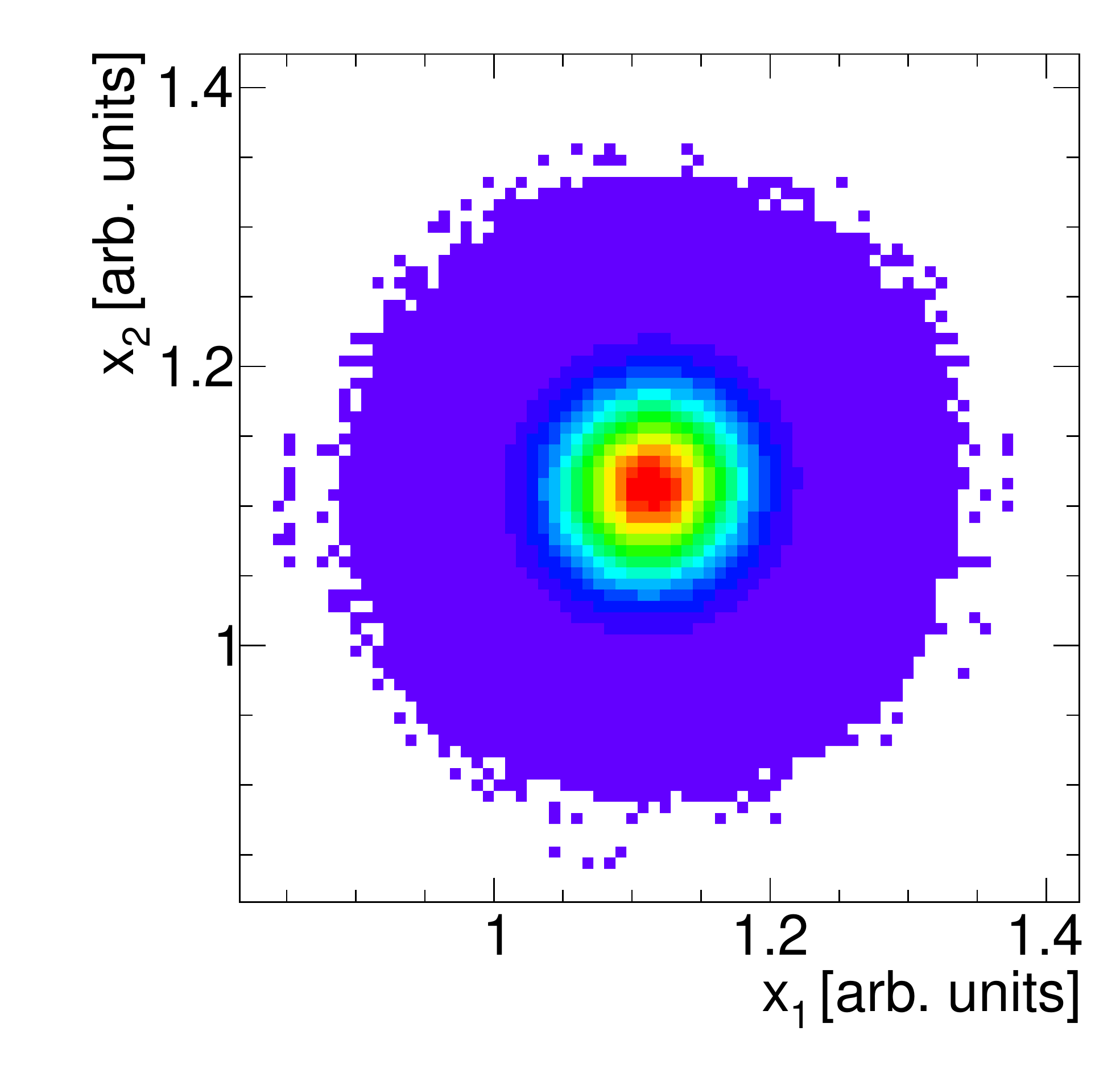}
\includegraphics[width=0.45\linewidth,clip=true]{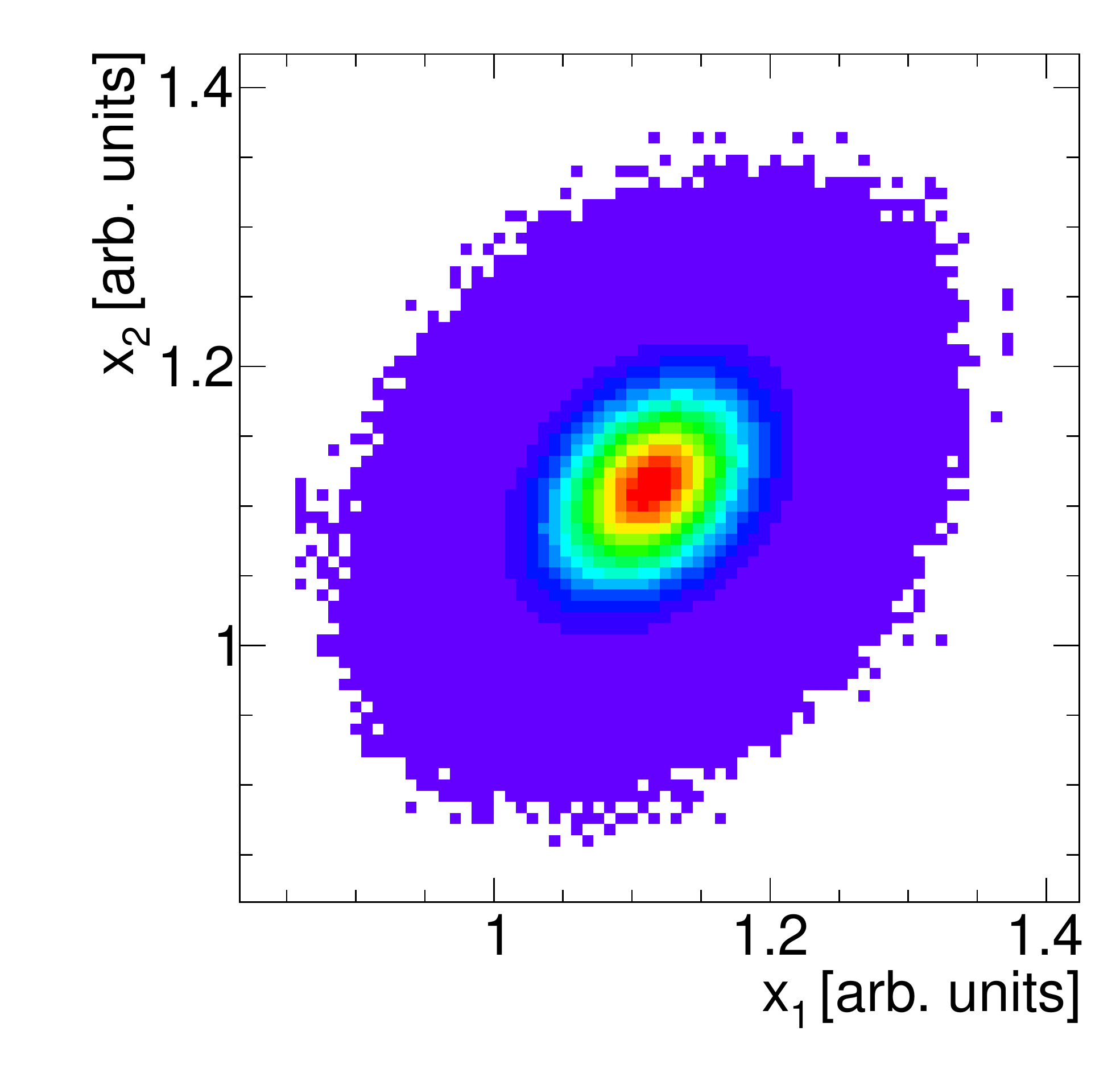}
\caption{(Color online)
 \textit{Left panel:}
Two-particle distribution $\rho(x_{1},x_{2})$ for pions with the correlation coefficient $R$ = 0.1.
\textit{Right panel:}
Similar to the left panel with the correlation coefficient $R$ = 0.5.
 }
\label{fig-2}
\end{figure}

\begin{figure}[htb]
\centering
 \includegraphics[width=0.6\linewidth,clip=true]{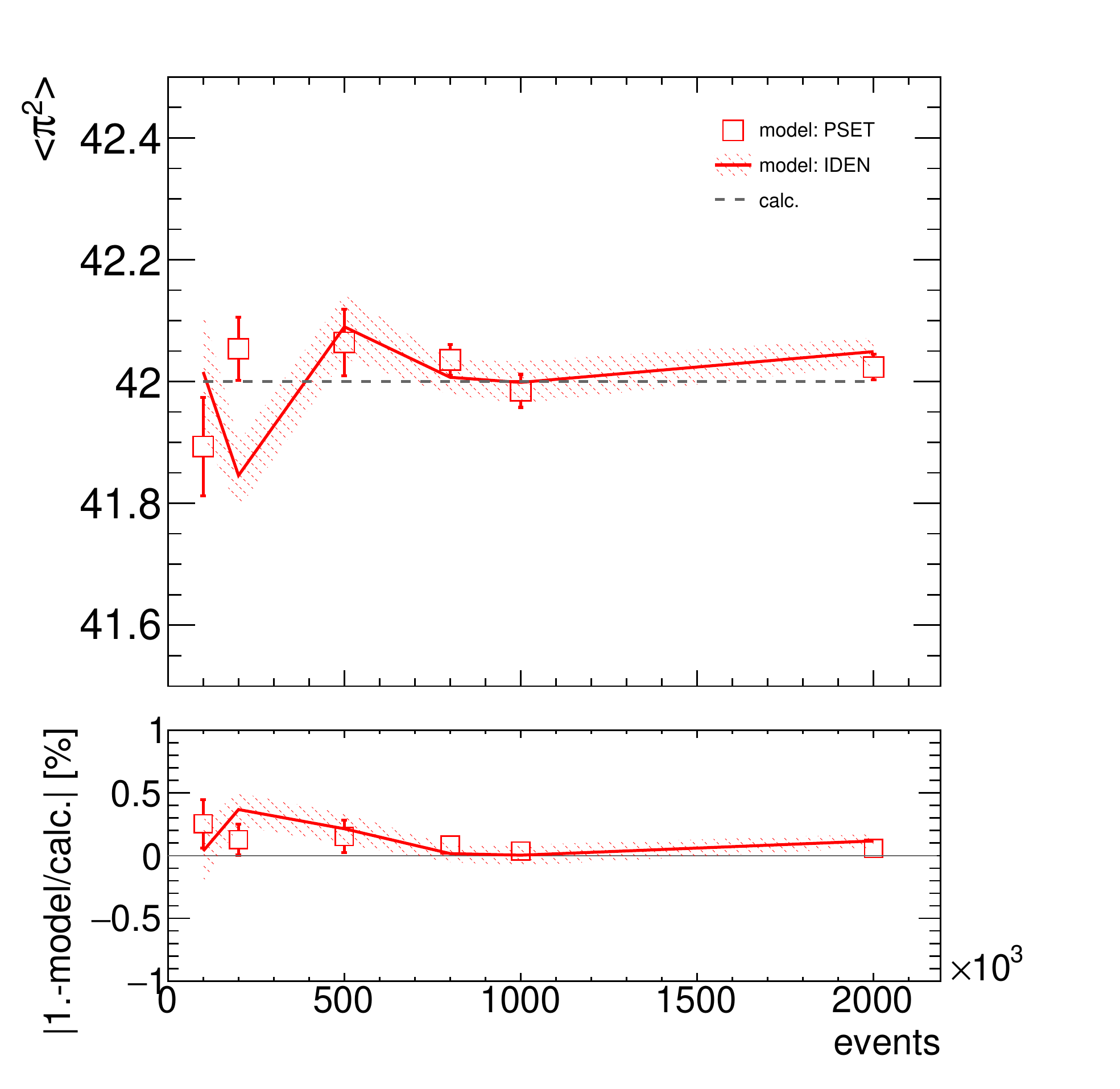}
\caption{(Color online)
 \textit{Upper panel:}
Results for the second moments of pions as computed with the PSET method (open boxes) and the Identity method (solid lines). The value of the correlation coefficient is taken to be $R$ = 0.1 (cf. left panel of Fig.~\ref{fig-2}). The analytical calculation is represented with the dashed line. 
\textit{Bottom panel:}
The ratios of the second moments of pions as computed with the PSET method to the theoretical baseline (open boxes). Similar ratios for the Identity method are presented with the solid lines. 
 }
\label{fig-3}
\end{figure}

\begin{figure}[htb]
\centering
 \includegraphics[width=0.6\linewidth,clip=true]{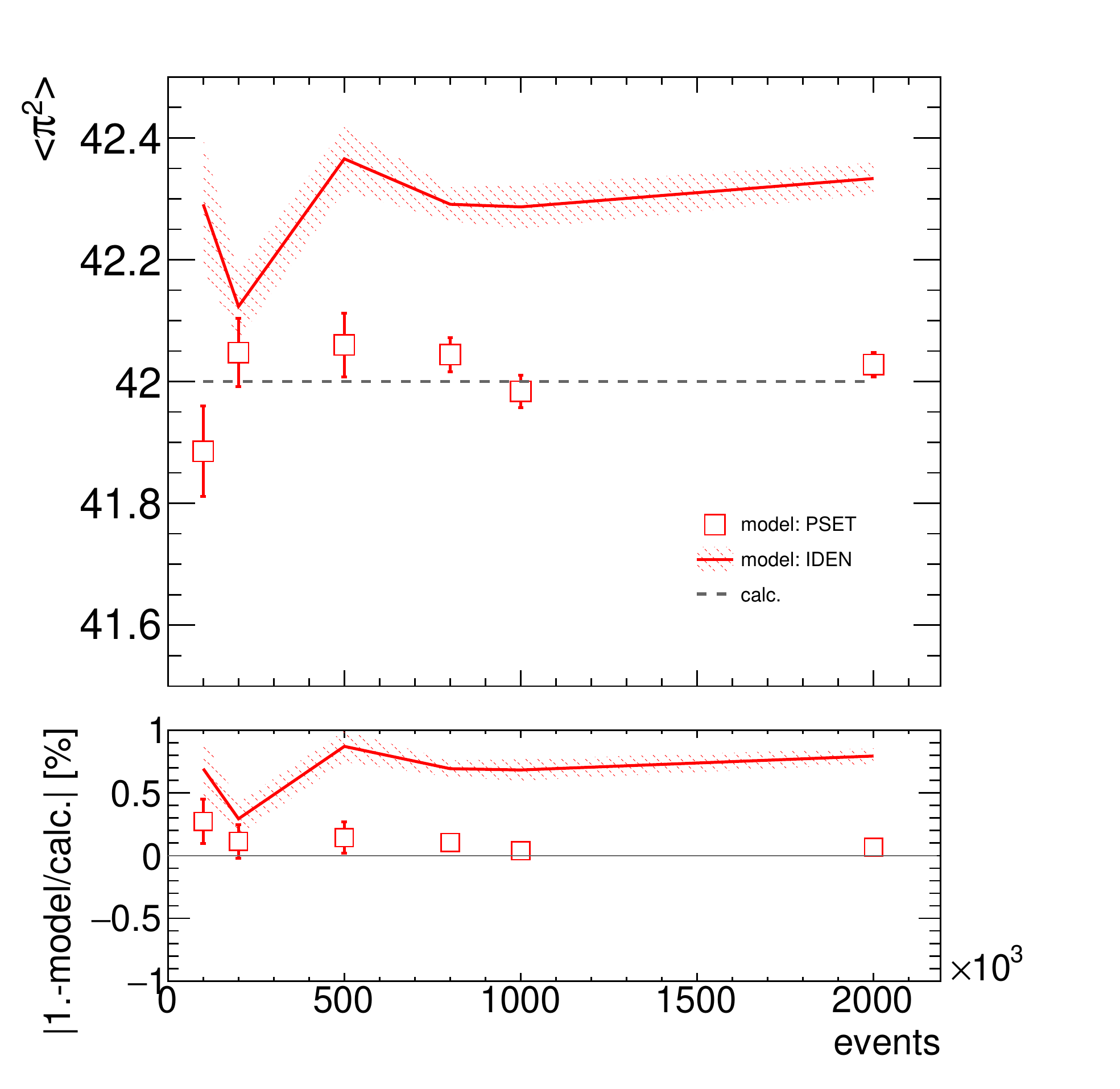}
\caption{(Color online)
 \textit{Upper panel:}
Results for the second moments of pions as computed with the PSET method (open boxes) and the Identity method (solid lines). The value of the correlation coefficient is taken to be $R$ = 0.5 (cf. right panel of Fig.~\ref{fig-2}). The analytical calculation is represented with the dashed line. 
\textit{Bottom panel:}
The ratios of the second moments of pions as computed with the PSET method to the theoretical baseline (open boxes). Similar ratios for the Identity method are presented with the solid lines. 
 }
\label{fig-4}
\end{figure}

In this section, based on simulated data, we provide results of the PSET Identification method and confront them with those obtained with the Identity method. Although the method is general and functions for unlimited number of particle species, for simplicity we consider the case of two particle types only. The simulation process consists of the following steps: (i) from independent Poisson distributions, with a given means of $\lambda_{\pi}=6$ and $\lambda_{K}=4$, we first randomly generate multiplicities of pions and kaons in each event; (ii) using the $\rho_{\pi}$ and $\rho_{K}$ distributions  functions, presented in the left panel of Fig.~\ref{fig-1} we generate the values of the particle identification variable $x$ corresponding to each particle species. In order to introduce correlations between pairs of $x$ quantities, we use the probability density function of the bi-variate normal distribution

\eq{ \label{eq:bivariate}
f(x_{1},x_{2}) = \frac{1}{2\pi\sqrt{|\Sigma|}}e^{-0.5\left(\mathbf{x}-\mathbf{\langle x\rangle}\right)^{T}\Sigma^{-1}\left(\mathbf{x}-\mathbf{\langle x\rangle}\right)},
}
where $|\Sigma|$ is the determinant of $\Sigma$. The column vectors $\mathbf{x}$, $\mathbf{\left<x\right>}$ and the covariance matrix $\Sigma$ are defined as

\eq{ \label{eq:bivariate2}
\mathbf{x} = \begin{pmatrix} x_{1} \\ x_{2} \end{pmatrix},
\mathbf{\langle x \rangle} = \begin{pmatrix} \langle x_{1}\rangle \\ \langle x_{2}\rangle \end{pmatrix},
\Sigma = \begin{pmatrix} \sigma_{x_{1}}^2 & R \sigma_{x_{1}}\sigma_{x_{2}}\\ R \sigma_{x_{1}}\sigma_{x_{2}} & \sigma_{x_{2}}^2 \end{pmatrix}.
}
The dimensionless  parameter $R$, referred to as the correlation coefficient, is introduced as

\eq{ \label{eq:bivariate3}
R =\frac{\left<(x_{1}-\left<x_{1}\right>)(x_{2}-\left<x_{2}\right>)\right>}{\sigma_{x_{1}}\sigma_{x_{2}}}.
}
We further note that the correlations between $x_{1}$ and $x_2$ are introduced only if they belong to the same particle, otherwise they are generated independently, i.e., the value of the correlation coefficient $R$ is set to zero in this case. 
Each generated $i$th event contains a set of quantities $X = \{x_1, x_2, ... , x_{N_i}\}$, where $N_i$ refers to the total number of pions and kaons in a given event. Next, we construct all possible two-particle pairs $(x_1,x_2)$ of the $x$ quantity inside a given event. The full set of these pairs in all events (cf. Eq.~\ref{eq:pairs.data})  generates an inclusive two particle mass distribution. 
From Eq.~\ref{eq:N2} we estimate second moments of pions. In doing so, we first fit the two-dimensional mass distribution (cf. the right panel of Fig.~\ref{fig-1}) and use Eq.~\ref{eq:rho2} to determine the mean numbers of pion-pion pairs $N_{\pi\pi}$. In a similar way we compute the second moments of the kaon multiplicity distribution. 
For clarity we present the results for pions only. In Fig.~\ref{fig-2} the two-particle distributions $\rho(x_{1},x_{2})$ are presented, where the left and right panels correspond to different values of the correlation coefficient: $R$ = 0.1 and 0.5 respectively. The corresponding reconstructed second moments are presented in the upper panels of Figs.~\ref{fig-3} and~\ref{fig-4}, where open boxes represent the results from the current study (PSET Identification method) while the  solid lines are obtained with the Identity method~\cite{Arslandok:2018pcu}. In the bottom panels of Figs.~\ref{fig-3} and~\ref{fig-4} the ratios of the second moments of pions to their theoretical values are presented.  Close inspection of Figs.~\ref{fig-3} and~\ref{fig-4} indicates that with the increasing correlations between $x_{1}$ and $x_{2}$ the Identity method deviates from the theoretical baseline, while the PSET Identification method, as expected, is protected against such correlations. We further note that, the amount of bias in the Identity method depends  on  mean multiplicities, number of involved particle species etc.

\section{Summary}
\label{sec:summary}
The paper presents a new method, the Particle-Set Identification method,
for reconstructing moments of multiplicity distribution of identified particles.
A PSET represents a set of
particles  which is constructed from particles created in a collision.
Mean multiplicities of particle sets of a given type are extracted from the measurements of
the multi-dimensional distribution of a particle "mass" variable.
This multi-dimensional distribution is used to calculate
moments of the joint multiplicity distributions of identified particles.

First, the PSET Identification  method  is introduced for the simple case of two particle types, addressing 
 first and second order moments.
Then a sketch of the generalization of the method is presented
for $k$-particle sets and moments of order $k\ge 3$ for the case of an arbitrary 
number of particle types. 

Finally, using a simple model we explicitly demonstrated that the PSET Identification  method is protected against possible correlations in the multi-dimensional distribution of the particle "mass" variables.

The PSET Identification method has a broader range of applicability than the Identity method
introduced by us previously to solve the problem of incomplete particle identification.
Particularly, it does not assume that measurements of particle "mass" for different
particles are independent. The issue of introducing momentum dependent pdfs within the PSET identification method is left for future studies.

\begin{acknowledgments} The authors acknowledge comments by Peter Seyboth and Maciek Lewicki. MG thanks Ola Snoch for motivating this work. 
The present work was partially supported by the Program of Fundamental Research of the Department of Physics and Astronomy of the National Academy of Sciences of Ukraine " Mathematics models of non-equilibrium process in open system" N 0120U100857,  
the Polish National Science Centre grants 2018/30/A/ST2/00226, 2016/21/D/ST2/01983, and the German Research Foundation grant GA1480\slash 2-2.
\end{acknowledgments}


\bibliographystyle{ieeetr}
\bibliography{references}
\end{document}